\documentclass{iaus}
\usepackage{graphicx}

\title[Developments in physics of massive stars] 
{Developments in physics of massive stars}

\author[Georges Meynet et al.]   
{Georges Meynet$^1$, Sylvia Ekstr\"om$^1$, , Andr\'e Maeder$^1$, Raphael Hirschi$^2$, Cyril Georgy$^1$, Coralie Beffa$^1$}
\affiliation{$^1$Observatory of Geneva University, Switzerland \\ email: {\tt georges.meynet@obs.unige.ch}\\[\affilskip]
$^2$EPSAM, University of Keele, UK \\ email: {\tt r.hirschi@epsam.keele.ac.uk}}

\pubyear{2008}
\volume{250}  
\pagerange{119--126}
\jname{Massive Stars as Cosmic Engines}
\editors{F. Bresolin, P.A. Crowther \& J. Puls, eds.}
\begin{document}

\maketitle

\begin{abstract}
New constraints on stellar models are provided by large surveys of massive stars, interferometric observations and asteroseismology. After a review of the main results so far obtained, we present new results from rotating models and discuss comparisons with observed features.
We conclude that rotation is a key feature of massive star physics.
 
\end{abstract}

\firstsection 

\section{Large surveys of massive stars}

\begin{figure}
\includegraphics[width=2.3in,height=2.6in,angle=-90]{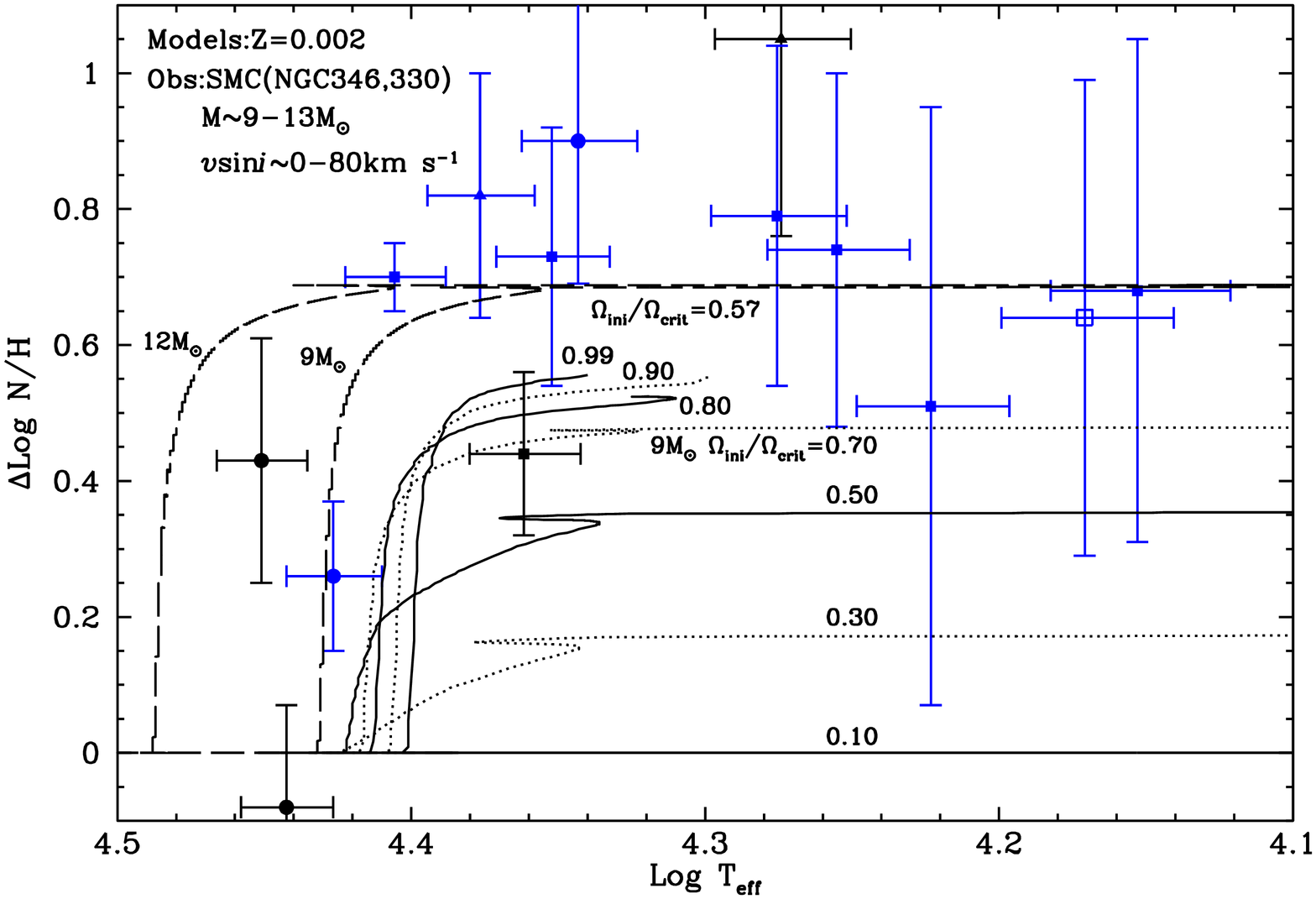}
\hfill
\includegraphics[width=2.3in,height=2.6in,angle=-90]{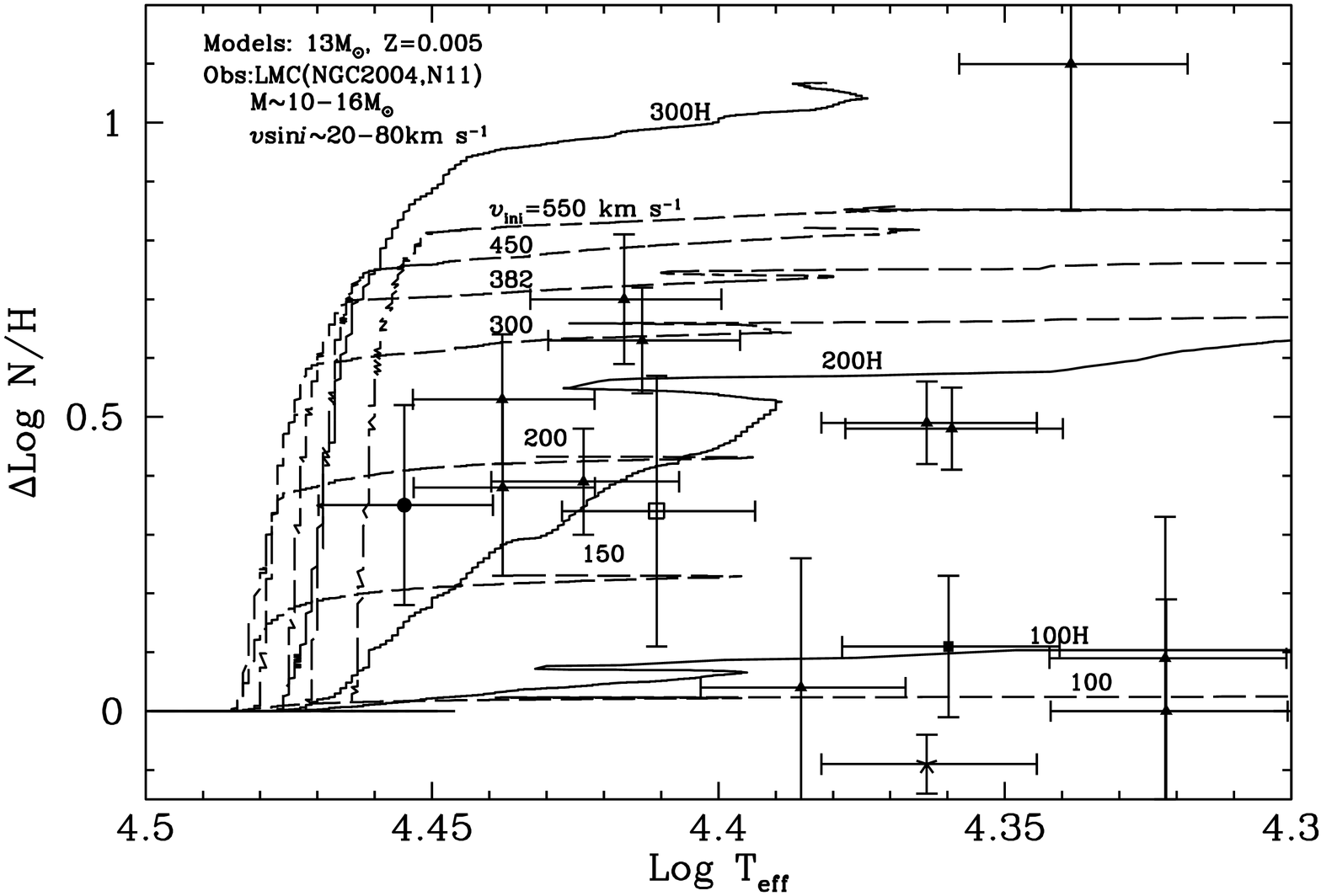}
\caption{
{\it Left panel}: 
Points with error bars are observations of SMC stars in the direction of NGC 346 by \cite[Hunter et al. (2007)]{Hunter_etal07} and of NGC 330 by  \cite[Trundle et al. (2007)]{Trundle_etal07}. Only stars with estimated masses between 9 and 13 M$_\odot$ have been considered. Continuous and dotted lines are the tracks computed by \cite[Ekstr\"om et al. (2008)]{Ekstrom_etal08}. Long-dashed lines are tracks computed by Ekstr\"om et al. (in preparation, see text).
{\it Right panel}: Observed points are for stars with estimated masses between 10 and 16M$_\odot$
taken from the same references as in the left panel. Long-dashed lines are rotating tracks computed for the present work with initial abundances as given by \cite[Hunter et al. (2007)]{Hunter_etal07} (see text). The continuous lines are models with rotation and magnetic fields (an H follows the initial equatorial velocity in that cases).
}\label{SMC920}
\end{figure}

Most of the developments in stellar physics arise from the necessity to
better reproduce observed features. For instance the observation of strong
nitrogen enrichments at the surface of main sequence OB stars indicates that
some extra mixing process is at work. Since massive stars are fast rotators and 
since rotation triggers many instabilities able to drive the transport of chemical species,
a lot of effort was put in order to properly modelize the effects of rotation. 
Results of these computations 
at their turn triggered large surveys of massive stars made possible thanks to
the advent of powerful multispectrographs. Among the most recent surveys let us mention
the following ones:

\begin{itemize}
\item \cite[Keller (2004)]{Keller04} presents measurements of the projected rotational velocities of a sample of 100 early B-type main-sequence stars in the Large Magellanic Cloud (LMC).
The sample is drawn from two sources: from the vicinity of the main-sequence turnoff of young clusters (ages 1-3 $\cdot$ 10$^{7}$ y) and from the general field.
\item \cite[Strom et al. (2005)]{Strom_etal05} measured the projected rotational velocities for 216 B0-B9 stars in the rich, dense $h$ and $\chi$ Persei double cluster and compared with the distribution of rotational velocities for a sample of field stars having comparable ages (~12-15 Myr) and masses (M~4-15 M$_\odot$).
\item \cite[Huang \& Gies (2006a)]{HuangGies06a} present projected rotational velocities for a total of 496 OB stars belonging to  young galactic clusters. Surface helium abundances are given in
\cite[Huang \& Gies (2006b)]{HuangGies06b}.
\item Martayan et al. (2006; 2007b) have measurements of $v\sin i$ of B and Be stars in the SMC (202 B, 131 Be) and in the LMC (121 B, 47 Be).
\item \cite[Wolff et al. (2007)]{Wolff_etal07} have measured projected rotational velocities for about 270 B-type stars belonging to galactic clusters or associations.
\item The VLT-Flames survey (still in progress) has for aim to analyze about 750 OB stars
observed in 7 fields centered on young clusters, 3 in the Milky Way (NGC 3293, 4755, 6611), 2 in the LMC (NGC 2004, N11) and 2 in the SMC ( NGC 330, NGC 346). The spectral classification and the radial velocities are discussed in Evans et al. (2005; 2006) together with some considerations on the
populations of binaries and of Be stars. Rotation velocities are discussed in Mokiem et al. (2006; 2007), \cite[Dufton et al. (2006)]{Dufton_etal06} and \cite[Hunter et al. (2008a)]
{Hunter_etal08a} (609 measurements published so far). First results concerning the surface chemical compositions are presented in 
Hunter et al. (2007; 2008b), \cite[Trundle et al. (2007)]{Trundle_etal07} (110 measurements of CNO and Mg Si published so far).
Helium surface enrichments at the surface of O and early B-type stars are discussed in 
Mokiem et al. (2006; 2007) (57 measurements published so far). 
\end{itemize}
\vskip 2mm
Concerning the distribution of {\bf rotational velocities}, the following results have been obtained:
\begin{itemize}
\item {\bf O-type stars in the SMC}: \cite[Mokiem et al. (2006)]{Mokiem_etal06} deduce the underlying $v$ distribution of the
unevolved SMC O-type stars. They obtain a mean velocity $v$ of about 150-180 km s$^{-1}$ and an effective half width of roughly 100-150 km s$^{-1}$.
\item {\bf OB-type stars in the MCs:} \cite[Hunter et al. (2008a)]{Hunter_etal08a} present the atmospheric parameters and the projected rotational velocities for approximately 400 O- and early B-type stars in the Magellanic Clouds. The observed $v \sin i$ distributions can be modeled by Gaussians with a peak at respectively 100 and 175 km s$^{-1}$ for the LMC and SMC and with a $1/e$ half width of 150 km s$^{-1}$ in both cases.
\item {\bf B- and Be-type stars in the MCs:} Martayan et al. (2006; 2007a) obtain mean $v \sin i$ of 161$\pm$20 km s$^{-1}$
and 155$\pm$20 km s$^{-1}$ for SMC B-type stars of respectively 2-5 (111 stars) and 5-10 M$_\odot$ (81 stars). Analogous stars in the LMC have mean projected rotational velocities of 144$\pm$13
and 119$\pm$11 km s$^{-1}$. Let us note that the average velocities of Be stars are much greater
(see the details in the above references). As an illustrative example, the mean $v \sin i$ for SMC Be stars in the mass ranges 2-5, 5-10, 10-12 and 12-18M$_\odot$ are respectively 277$\pm$34 (14 stars), 297$\pm$25 (81 stars), 335$\pm$20 (13 stars) and  336$\pm$40 (14 stars). 
\item {\bf B-type stars in the Galaxy:} \cite[Huang \& Gies (2006a)]{HuangGies06a} present projected rotational velocities for 496 OB stars belonging to 19 young galactic clusters with estimated ages between 6 and 73 Myr.
Mean $v\sin i$ values of 139, 154 and 151 km s$^{-1}$ have been obtained for groups of
O9.5-B1.5, B1.5-B5.0 and B5.0-B9.0 type stars. These authors derived the underlying probability distribution for the equatorial velocities $v$ and obtained a peak at 200 km s$^{-1}$.
\cite[Dufton et al. (2006)]{Dufton_etal06} obtain a peak of $v$ at 250 km s$^{-1}$ with a full-width-half-maximum of approximately 180 km s$^{-1}$ for the unevolved targets in the galactic clusters NGC 3293 and 4755.
\end{itemize}
\vskip 2mm
Some authors have discussed the variation of the velocity as a function of the age of the stars.
The main results are the following: 
\begin{itemize}
\item {\bf O-type stars in the SMC:} According to \cite[Mokiem et al. (2006)]{Mokiem_etal06} who analyses O-type stars in the SMC, the observed distribution of $v\sin i$ for evolved stars (luminosity classes I-II) contains relatively fewer fast rotators and slow rotators compared to the distribution for unevolved stars
(luminosity classes IV-V). A similar trend is obtained for galactic stars. These authors suggest that when the star evolves, it
undergoes a spin down due to an increased radius, a loss of angular momentum through stellar winds.
This may explain the smaller proportion of fast rotators. The smaller proportion of slow rotators
might be due to excess of turbulent broadening among evolved stars.
\item {\bf OB-type stars in the MCs:} \cite[Hunter et al. (2008a)]{Hunter_etal08a} from a sample of 400 O- and early B-type stars in the Magellanic Clouds also find that supergiants are the slowest rotators in the sample, typically having rotational velocities less than 80 km s$^{-1}$. 
\item {\bf OB-type stars in the Galaxy:} \cite[Huang \& Gies (2006b)]{HuangGies06b} show that all OB stars of their sample experience a spin-down during the MS phase. A few relatively fast rotators are found near the TAMS. According to these authors, these stars may be spun up by a short contraction phase or by mass transfer in a close binary.
\item \cite[Wolff et al. (2007)]{Wolff_etal07} find that independent of environment, the rotation rates for stars in the mass range 6-12 M$_\odot$ do not change by more than 0.1 dex over ages between about 1 and 15 Myr. 
\end{itemize}
\vskip 2mm
Rotation may also depend on the mass and on the metallicity  as is deduced from the results below:
\begin{itemize}
\item {\bf Higher masses, lower velocities:} According to \cite[Hunter et al. (2008a)]{Hunter_etal08a} there is some evidences that the most massive objects rotate slower than their less massive counterparts.
\cite[Dufton et al. (2006)]{Dufton_etal06} find that the mean rotational velocity of stars which have strong winds is lower than that of the lower mass stars.
\item {\bf Lower $Z$, higher velocities:} \cite[Mokiem et al. (2006)]{Mokiem_etal06} find that among O-type stars, the distribution for unevolved SMC objects shows a relative excess of stars
rotating with projected velocities between 120 and 190 km s$^{-1}$ compared to analogous velocity distributions in the Galaxy. This can be interpreted as a decrease of angular momentum loss by stellar winds in lower metallicity environments. \cite[Hunter et al. (2008a)]{Hunter_etal08a} also obtain that SMC metallicity stars rotate on average faster than galactic ones (mainly field objects). No difference is found between galactic and LMC stars. 
\cite[Martayan et al. (2007a)]{Maratayan_etal07a} find that, for B and Be stars, the lower the metallicity, the higher the rotational velocities.
\end{itemize}

\vskip 2mm
Many authors find that rotational velocities of OB stars in clusters are greater than
those of stars belonging to less dense systems like stellar associations or the field: for instance
\begin{itemize}
\item \cite[Keller (2004)]{Keller04} obtains that the mean $v\sin i$ of early B-type stars in clusters with ages between 1-3$\cdot$10$^7$ years of the LMC is 146 km s$^{-1}$, while it is 112 km s$^{-1}$
for analogous field stars. A same trend has been found for galactic stars although with lower
values for both the clusters (116 km s$^{-1}$) and the field (85 km s$^{-1}$).
\item \cite[Strom et al. (2005)]{Strom_etal05} find that B-type stars members of $h$ and $\chi$ Per
(age between 12 and 15 Myr)
have mean $v\sin i$ higher than analogous field stars. The difference between these two means
depends on the evolutionary stage. For less evolved stars (4-5 M$_\odot$), the mean projected
velocity is 183 km s$^{-1}$ for cluster stars and 92 km s$^{-1}$ for field stars, for
somewhat more evolved stars (5-9 M$_\odot$), the cluster and field mean $v \sin i$ are
145 and 93 km s$^{-1}$, while 
for stars approaching the end of the Main-Sequence phase (9-15 M$_\odot$), one has respectively
104 and 83 km s$^{-1}$. 
\item \cite[Dufton et al. (2006)]{Dufton_etal06} find that the projected velocities in the galactic clusters NGC 3293 and 4755 are systematically larger than those for the field. 
\cite[Huang \& Gies (2006a)]{HuangGies06a} find from their study of galactic stars that
there are more fast rotators among the B cluster stars than in the case of the field stars.
The mean projected rotational velocities are 148$\pm$4 km s$^{-1}$ and 113$\pm$3 km s$^{-1}$ for the cluster and field stars respectively.
\item \cite[Wolff et al. (2007)]{Wolff_etal07} obtain that 
stars formed in high-density regions lack the cohort of slow rotators that dominate the low-density regions and young field stars. 
\end{itemize}

\begin{figure}
\includegraphics[width=2.3in,height=2.6in,angle=-90]{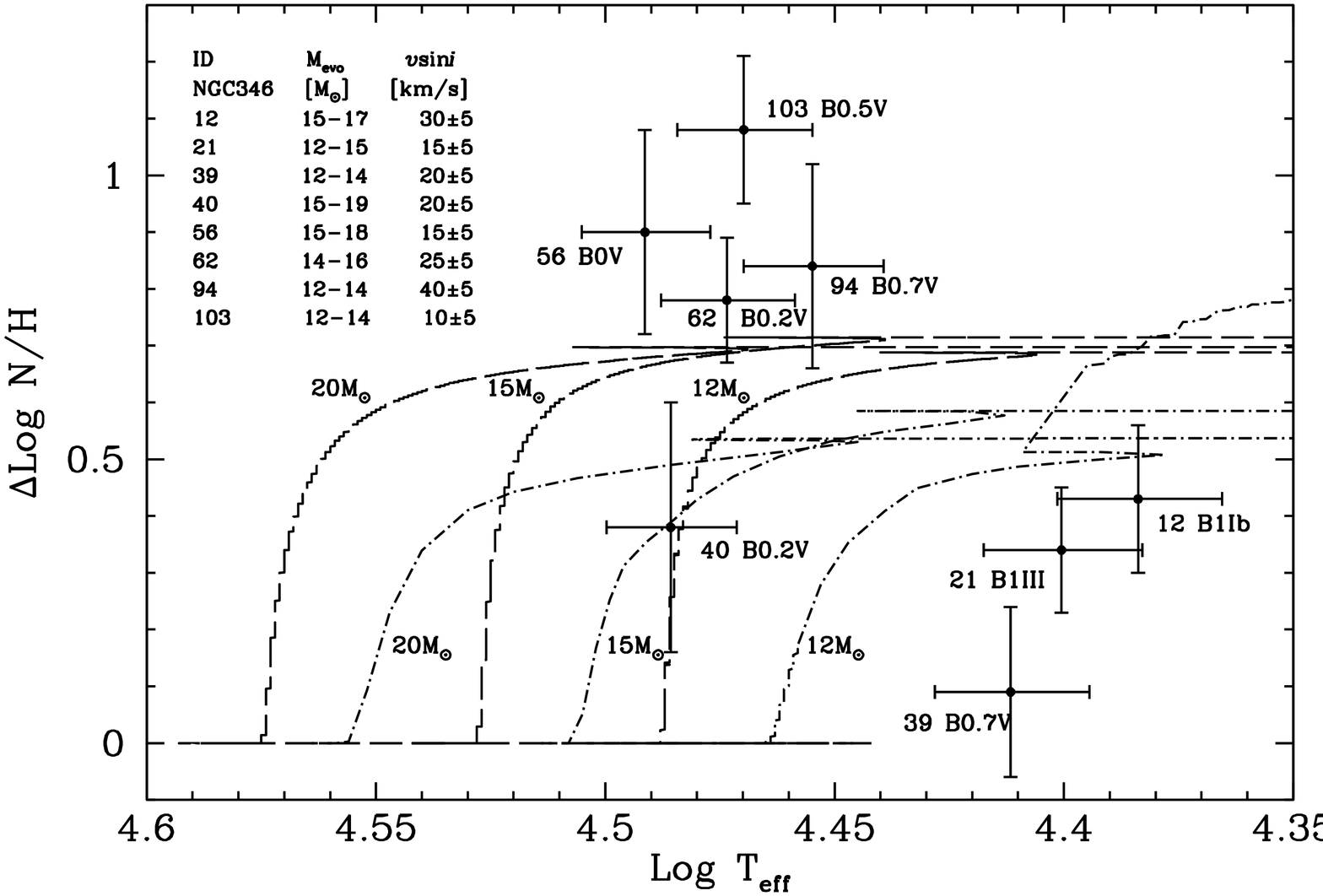}
\hfill
\includegraphics[width=2.3in,height=2.6in,angle=-90]{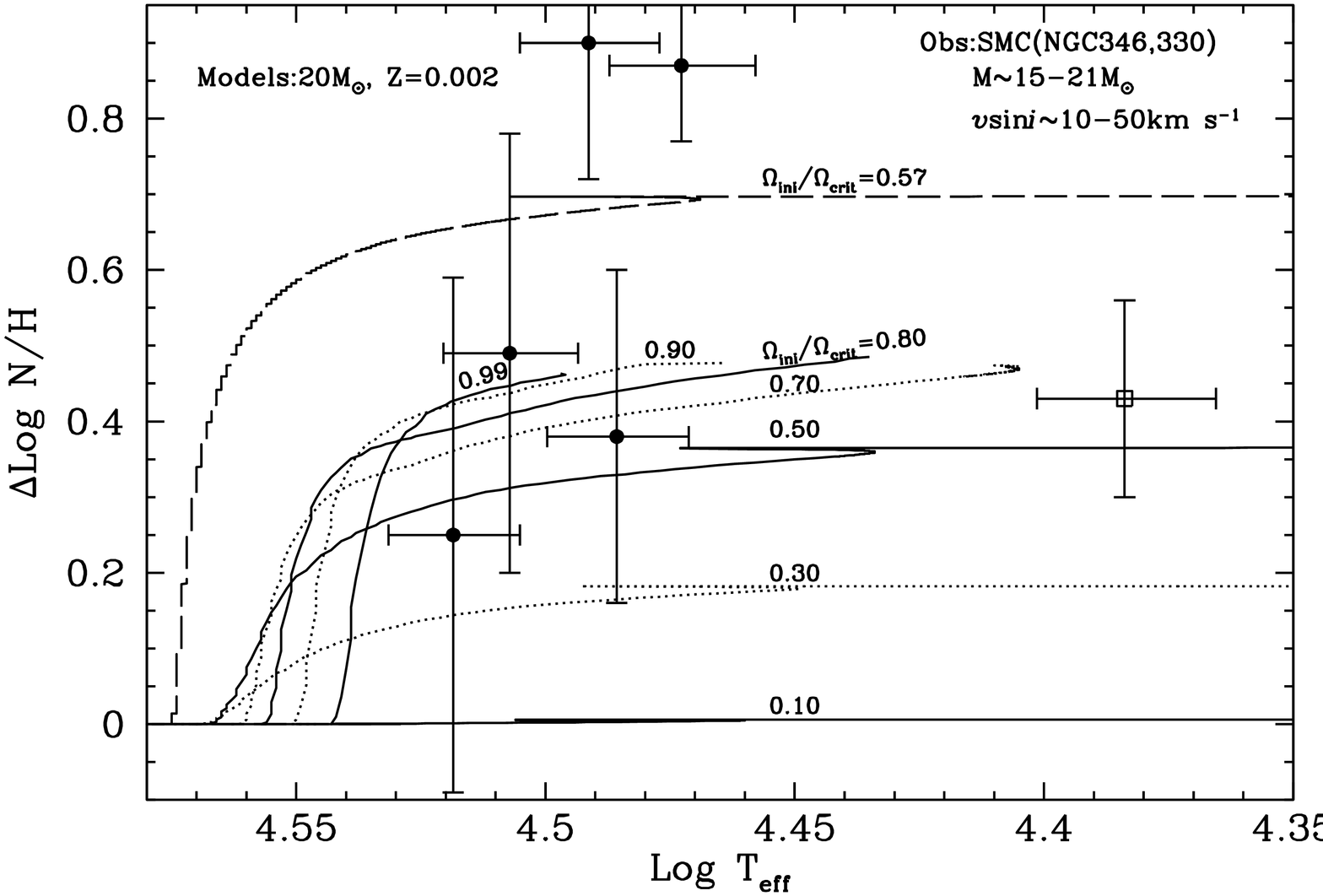}
\caption{{\it Left panel}: Observed points are from \cite[Hunter et al. (2007)]{Hunter_etal07}.
The long-dashed tracks are from  the rotating models of Ekst\"om et al. (in preparation, see text) with $v_{\rm ini}/v_{\rm crit}=0.4$, the dot-short dashed tracks are rotating models with
$v_{\rm ini}=300$ km s$^{-1}$ from \cite[Maeder \& Meynet (2001)]{MMVII}.
The models of Ekst\"om et al. have initial values of $\Omega_{\rm ini}/\Omega_{\rm crit}=0.57$,
while the models of \cite[Maeder \& Meynet (2001)]{MMVII} have 
$\Omega_{\rm ini}/\Omega_{\rm crit}=0.64$, 0.61 and 0.58 for respectively the 12, 15 and 20M$_\odot$.
{\it Right panel}: Points with error bars are observations of SMC stars in the direction of NGC 346 by \cite[Hunter et al. (2007)]{Hunter_etal07} and of NGC 330 by  \cite[Trundle et al. (2007)]{Trundle_etal07}. Only stars with estimated masses between 15 and 21 M$_\odot$ have been considered. The star NGC330-003 has not been plotted being well outside the range of the other stars, it would lie at the
position given by log T$_{\rm eff}$=4.235 and $\Delta$log N/H=1.19.
Continuous and dotted lines are the tracks computed by \cite[Ekstr\"om et al. (2008)]{Ekstrom_etal08}. Long-dashed lines are tracks computed by Ekst\"om et al. (in preparation, see text).
}\label{LMC13SMC15}
\end{figure}
\vskip 2mm
The Be stars are stars surrounded by an expanding equatorial disks problably
produced by the concomitant effects of both fast rotation and pulsation. These objects are wonderful
laboratories to study the effects of extreme rotation. Many new results concerning them have been
obtained:
\begin{itemize}
\item Objects with Be phenomena are the fastest rotators in the sample studied by \cite[Hunter et al. (2008a)]{Hunter_etal08a} (400 OB stars in the MCs). This trend is confirmed by
Martayan et al. (2006; 2007a) who obtain that Be stars rotate faster than B stars whatever the metallicty is (see above). 
\item These last authors obtain that Be stars with masses below about 12M$_\odot$ are mainly observed in the second part of the MS whatever the metallicity. The more massive stars are mainly in the first part of the MS in the MW, while in the Magellanic Clouds, they are all in the second part of the MS.
\item \cite[Martayan et al. (2007b)]{Maratayan_etal07b} find 13 Be stars among the sample of Be SMC stars with short-term periodicity and 9 of them are multi-periodic pulsators. The detected periods fall in the range of slowly pulsating B-type  stars modes (from 0.4 to 1.60 days).
\item \cite[Maeder et al. (1999)]{Maeder_etal99} and \cite[Wisniewski \& Bjorkman (2006)]{WisBjork06}
find that the fraction of Be stars with respect to the total number of B and Be stars
in clusters with ages (in years)  between 7.0 and 7.4 (in logarithm) increases when the metallicity decreases. This fraction passes from about 10\% at solar metallicity to about 35\% at the SMC metallicity. These significant proportions imply that Be stars may drastically affect the mean $v \sin i$ obtained for a given population of B-type stars. 
\end{itemize}
\vskip 2mm
{\bf Surface enrichments in helium and nitrogen} have been observed with the following
main trends (see also Maeder in these proceedings):
\begin{itemize}
\item {\bf He in O-type stars in the SMC:} In the SMC, for 31 O-type stars, \cite[Mokiem et al. (2006)]{Mokiem_etal06}
find values of $y=n_{\rm He}/(n_{\rm H}+n_{\rm He})$ between 0.09 and 0.24, where $n_{\rm He}$ is the density number of helium and $n_{\rm H}$ of hydrogen . Note that $n_{\rm He}/(n_{\rm H}+n_{\rm He})=Y/(Y+4X)$, where $Y$ and $X$ are respectively the mass fraction of helium and of hydrogen. 
Setting $Y\sim 1-X$, one obtains values for $Y=4y/(1+3y)$ between 0.28 and 0.56
(here 0.28 would correspond to y=0.09 {\it i.e.} to the initial helium mass fraction).
These authors conclude that while rotation can qualitatively account for such enrichments, the observed enrichments are in many cases much stronger than those predicted by the models.
\item {\bf He in O-type stars in the LMC:} In the LMC, for 28 O-type stars, $y$ values between about 0.09 and 0.28 
are obtained by \cite[Mokiem et al. (2007)]{Mokiem_etal07}, {\it i.e.} helium mass fractions between
0.28 and 0.61.
\item {\bf He in OB-type stars in the MW:} 
\cite[Huang \& Gies (2006b)]{HuangGies06b} determine He abundances for OB galactic stars.
In their high mass range ($8.5$M$_\odot < M < 16$M$_\odot$), the He enrichment progresses through the main sequence and is greater among the faster rotators\footnote{These authors also found many
helium peculiar stars (He-weak and He-strong). These stars were not used to study the process of He-enrichment.}. On average He abundance increases of 23\% $\pm$13\% between ZAMS and TAMS.
These authors also obtain that He enrichments are higher for higher $v \sin i$ values.
\cite[Lyubimkov et al. (2004)]{Lyubimkov_etal04} find a ZAMS to TAMS increase in He abundance of 26\% for stars in the mass range 4-11M$_\odot$ and 67\% for more massive stars in the range
12-19M$_\odot$.
\item {\bf N in O-type stars in the SMC:}
\cite[Heap et al. (2006)]{Heap_etal06} study a sample of 18 O-type stars in the direction of the SMC cluster NGC 346. The surface of about 80\% of the stars is moderately to strongly enriched in nitrogen, while showing the original helium, carbon and oxygen abundances.  
\item {\bf N in OB-type stars in the MCs:}
At the present time, the published values of nitrogen surface abundances from the VLT-Flames survey concern stars with relatively low $v\sin i$ (see Trundle et al. 2007 and Hunter et al. 2007\footnote{In \cite[Hunter et al. (2008b)]{Hunter_etal08b}
nitrogen surface abundances for stars with high $v\sin i$ are discussed but
no detailed tables with the individual measurements are provided.}). Some of these observations for SMC and LMC stars are shown in Figs.~\ref{SMC920} and \ref{LMC13SMC15}. A large spread of abundances is found  spanning a range between 0 and 1.19 dex of N/H enhancements for masses between 9 and 21M$_\odot$.
\item {\bf N in B-type stars in the MW:}
The galactic stars do not seem to present large spread of nitrogen abundances like that seen for Magellanic Cloud stars. 
\cite[Trundle et al. (2007)]{Trundle_etal07} note that if the galactic stars underwent the same degree of enrichment as the LMC and SMC stars (in absolute value), this would amount to only a factor of two or 0.3 dex enhancement  in the Galaxy. Such enhancements are similar to the degree
of uncertainties in their measurements. 
\end{itemize}

\begin{figure}
\includegraphics[width=2.0in,height=2.4in,angle=-90]{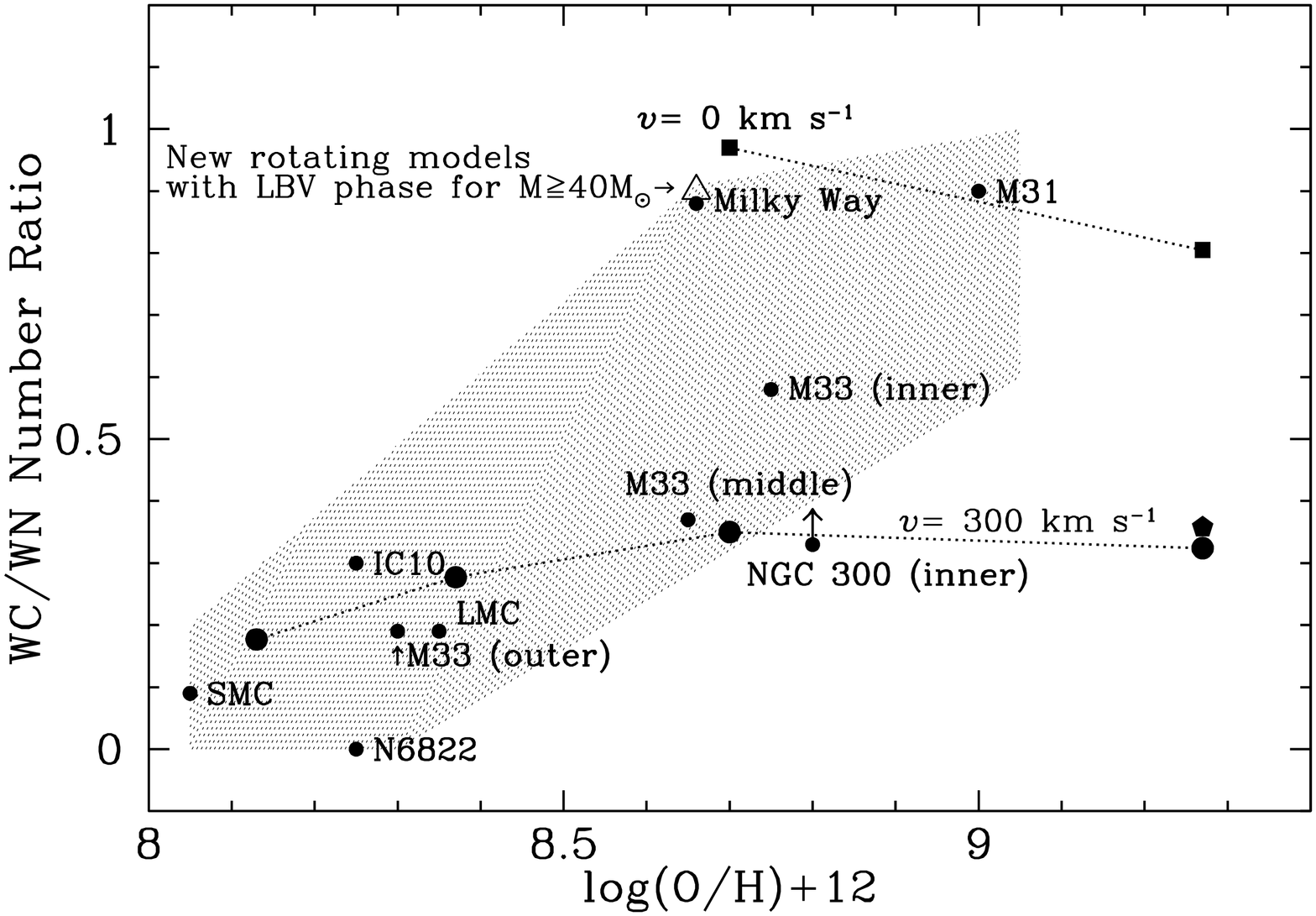}
\hfill
\hspace{1cm}
\includegraphics[width=2.0in,height=2.4in,angle=-90]{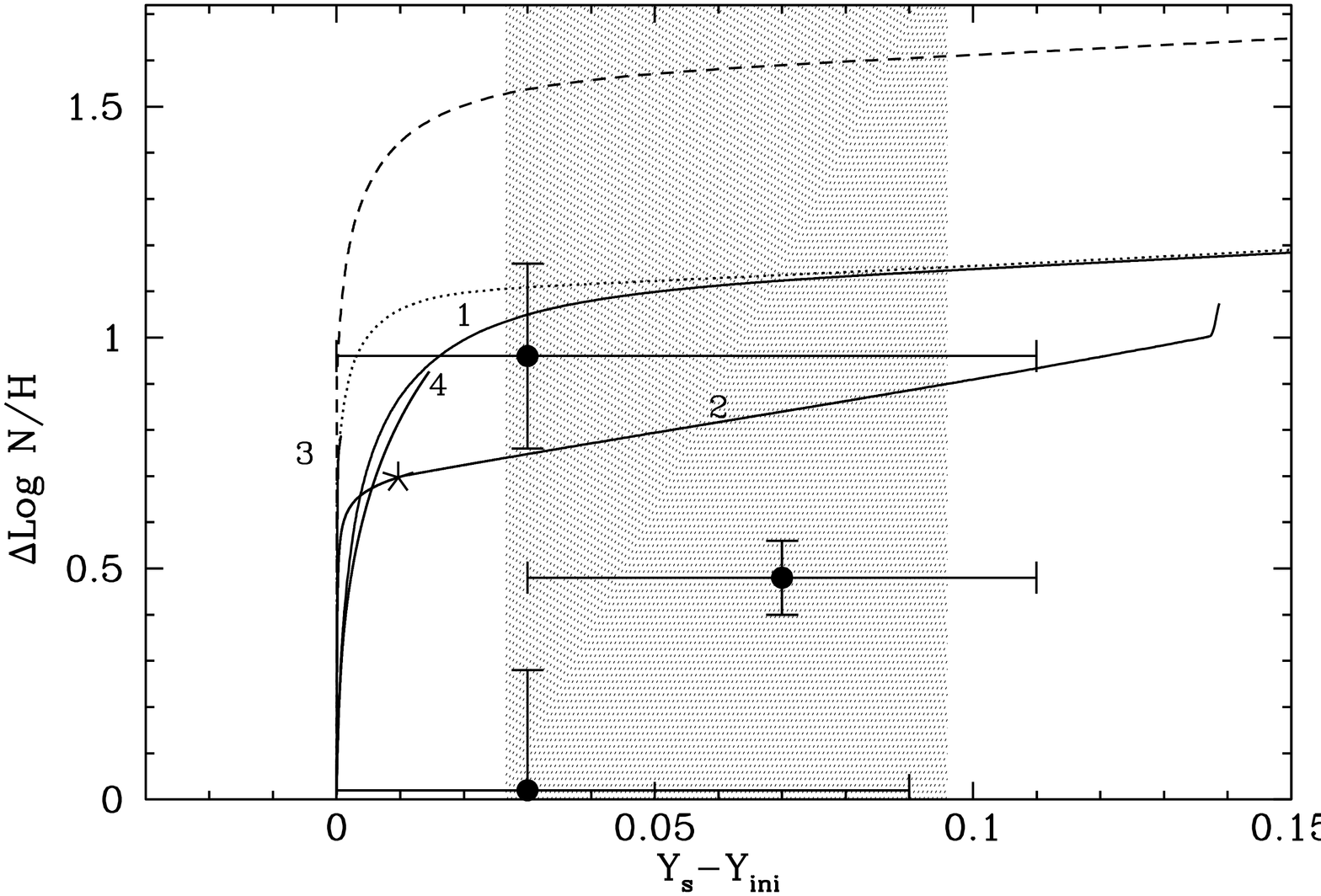}
\caption{{\it Left panel:} Variation of the number ratios of WN to WC stars as a function of metallicity. The grey area
encompasses the observed ratios. Individual measures are indicated by
black circles labeled with the name of the Galaxy (see references in Meynet \& Maeder 2005). Solar (O/H) value
is taken from \cite[Asplund et al. (2005)]{AS05}, the (O/H) values for the SMC and LMC are taken from \cite[Hunter et al. (2007)]{Hunter_etal07}.
The dotted lines show the predictions of the rotating and non--rotating stellar models
of \cite[Meynet \& Maeder (2005)]{MMXI}. The black pentagon shows the ratio predicted by Z=0.040 models computed
with the metallicity dependence of the mass loss rates during the WR phase. The open triangle shows the WC/WN ratio obtained from the new rotating models (see text).
{\it Right panel:} 
Continuous lines show the evolution of the nitrogen surface enrichment as a function of the helium surface enrichment. Model 1 is the fast rotating 60M$_\odot$ model computed by \cite[Meynet \& Maeder (2007)]{MMGRB} ($Z=0.002$, $v_{\rm ini}=$523 km s$^{-1}$.
This model follows a nearly ``homogeneous'' evolution. Model 2 is a 20M$_\odot$ model for $Z=0.002$ with an initial velocity on the ZAMS of 304 km s$^{-1}$, Models 3 and 4 are 13M$_\odot$ for $Z=0.005$ with an initial velocity on the ZAMS of 382 km s$^{-1}$ without and with a magnetic field.
The dotted line shows the typical evolution at the centre for the initial composition used
to compute the $Z=0.002$ models, the dashed line shows the evolution at the centre when the initial composition of the LMC as given by \cite[Hunter et al. (2007)]{Hunter_etal07} are adopted.
For model 1 the whole track shown occurs during the MS phase, for model 2, the star indicates the end
of the MS phase, for tracks 3 and 4 the track ends at the end of the MS phase. The grey area
shows the He enrichment obtained by \cite[Huang \& Gies (2006b)]{HuangGies06b} at the end of the MS.
The three dots correspond to three stars observed in the direction of the LMC cluster N11 
by \cite[Mokiem et al. (2007)]{Mokiem_etal07} and \cite[Hunter et al. (2007)]{Hunter_etal07} for which He enrichments have been found 
From top to bottom, one has the stars N11-008 (B0.7Ia,$~\sim 31$M$_\odot$), N11-072 (B0.2III,$~\sim 14$M$_\odot$) and N11-042 (B0III,$~\sim 25$M$_\odot$).
}\label{tma}
\end{figure}

\vskip 2mm
The problem of the mass discrepancy (see Herrero et al 1992), 
{\it i.e.} of the difference obtained between spectroscopic and evolutionary masses, has been
significantly reduced thanks to improvements brought to stellar atmosphere models. However
still some discrepancies are reported in general linked with strong helium surface enrichments:
\begin{itemize}
\item {\bf In SMC:} \cite[Mokiem et al. (2006)]{Mokiem_etal06} find a mild mass discrepancy for stars with 
spectroscopic masses inferior to about 20M$_\odot$, which correlates with the surface helium
abundance. These authors find that the discrepancies are consistent with the predictions of chemically homogeneous evolution. Most of the stars observed by \cite[Heap et al. (2006)]{Heap_etal06} exhibit the mass discrepancy problem although no surface He enrichment.
\item {\bf In LMC:} \cite[Mokiem et al. (2007)]{Mokiem_etal07} from the analysis of O-type stars in the LMC find that bright giants and supergiants do not show any mass discrepancy, regardless of the surface helium abundance. In contrast they find that the spectroscopically determined masses of the dwarfs and giants are systematically smaller than those derived from non-rotating evolutionary tracks. All dwarfs and giants having $y > 0.11$ ($Y >0.33$) show this mass discrepancy.
\end{itemize}

\vskip 2mm
Close binary stars have a very different evolution from single stars. Some clues pointing to these
differences are indicated below:
\begin{itemize}
\item {\bf Rotational velocities:} \cite[Hunter et al. (2008a)]{Hunter_etal08a} find that binaries tend to rotate slower than single objects. This results however may be somehow biased by the fact that binarity is easier to detect at low projected rotational velocities.
\cite[Huang \& Gies (2006a)]{HuangGies06a} on the contrary find that their binary candidates
({\it i.e.} those stars with a difference of radial velocity larger than 50 km s$^{-1}$), have a mean
$v \sin i$ of 173$\pm$15 km s$^{-1}$, higher than 144$\pm$5 km s$^{-1}$ the mean for the remaining  constant radial-velocity stars. 
\cite[Huang \& Gies (2006b)]{HuangGies06b} find that close binaries generally experience a significant spin-down around the stage where polar gravity is equal to 3.9 in logarithm. According to these authors, that is probably the result of tidal interaction and orbital synchronization.
\item {\bf Surface abundances:} According to \cite[Hunter et al. (2007)]{Hunter_etal07} main-sequence binary objects have close to baseline nitrogen surface abundances. These systems thus do not present apparent signs of 
extra-mixing. In contrast several evolved binary objects have high nitrogen enhancements. These
abundances are similar to those observed in apparently single stars. Thus it appears difficult to
discriminate among the possible causes of these enrichments, {\it i.e.} between extra-mixing operating in single stars and mass-transfer events in close binary systems.
The same result has been obtained by \cite[Trundle et al. (2007)]{Trundle_etal07}.
\end{itemize}

The above listed results are of course not exhaustive and many more might have been cited as the variation of the effective temperature scale for OB stars at different metallicities or the variation of the
mass loss with the metallicity. Concerning this last point let us just mention that \cite[Mokiem et al. (2006)]{Mokiem_etal06} find that for stars with log$L$/L$_\odot$ superior to about 5.4, the wind strengths are in excellent agreement with the theoretical predictions of \cite[Vink et al. (2001)]{Vink_etal01}.

Obviously the evolution of rotation and of the surface enrichments depend on the mass, the age, the metallicity, the environment (field/clusters), binarity and probably on other factors as magnetic fields. Thus the task to disentangle observationally all these different effects is very challenging and requires in addition to large surveys detailed observations of a few systems for which many precise measurements can be performed.   

\section{Interferometry and asteroseismology}

In addition to the above large surveys, there are at least two observational techniques which now begin to be applied to massive stars. The first one is interferometry. This technique allows to measure the shape of stars, the variation with the colatitude of the effective temperature, the inclination angle, the shape of the stellar wind as well as some characteristics of stellar disks.
Among recent very interesting results let us mention:
\begin{itemize}
\item \cite[Meilland et al. (2007b)]{Meilland_etal07b}
present the global geometry of the disk around the Be star $\alpha$ Arae. The 
global geometry of the disk is compatible with a thin keplerian disk and polar enhanced winds
(see also Kervella \& Domiciano de Souza 2006).
They also obtain that $\alpha$ Arae is rotating very close to its critical rotation
(for results on the Be star $\kappa$ CMa see Meilland et al. 2007a).
\item \cite[Domiciano de Souza et al. (2007)]{Domiciano_etal07} have performed 
the first high spatial and spectral resolution observations of the circumstellar envelope  of a B[e] supergiant (CPD-57°2874).
\item \cite[Millour et al. (2007)]{Millour_etal07}, using 
AMBER/VLTI observations of the Wolf-Rayet and O (WR+O) star binary system $\gamma$2 Velorum
deduce that the binary system lies at a distance of 368+38-13 pc, in agreement with recent spectrophotometric estimates, but significantly larger than the Hipparcos value of 258+41-31 pc.
\item \cite[Weigelt et al. (2007)]{Weigelt_etal07} have made the
first NIR spectro-interferometry of the LBV $\eta$ Carinae.
Their observations support theoretical models of anisotropic winds from fast-rotating, luminous hot stars with enhanced high-velocity mass loss near the polar regions.
\end{itemize}

The second technique, the asteroseismology, provides new insights on massive star interiors (see the paper by C. Aerts in these proceedings). At the moment, data for five B-type stars 
with masses between 8 and 14 M$_\odot$ have been obtained. The core overshoot parameter expressed
in units of pressure scale height has been found to be of the order of  0.20 (two stars are compatible with that value, two with 0.10 and one with 0.44). For three stars the values of the ratio
of core to envelope angular velocity have been obtained. The values are 5, 3.6 and 1 (solid body rotation).

In the next sections we discuss a few results recently obtained from massive star rotating models.

\begin{figure}
\includegraphics[scale=0.32,angle=0]{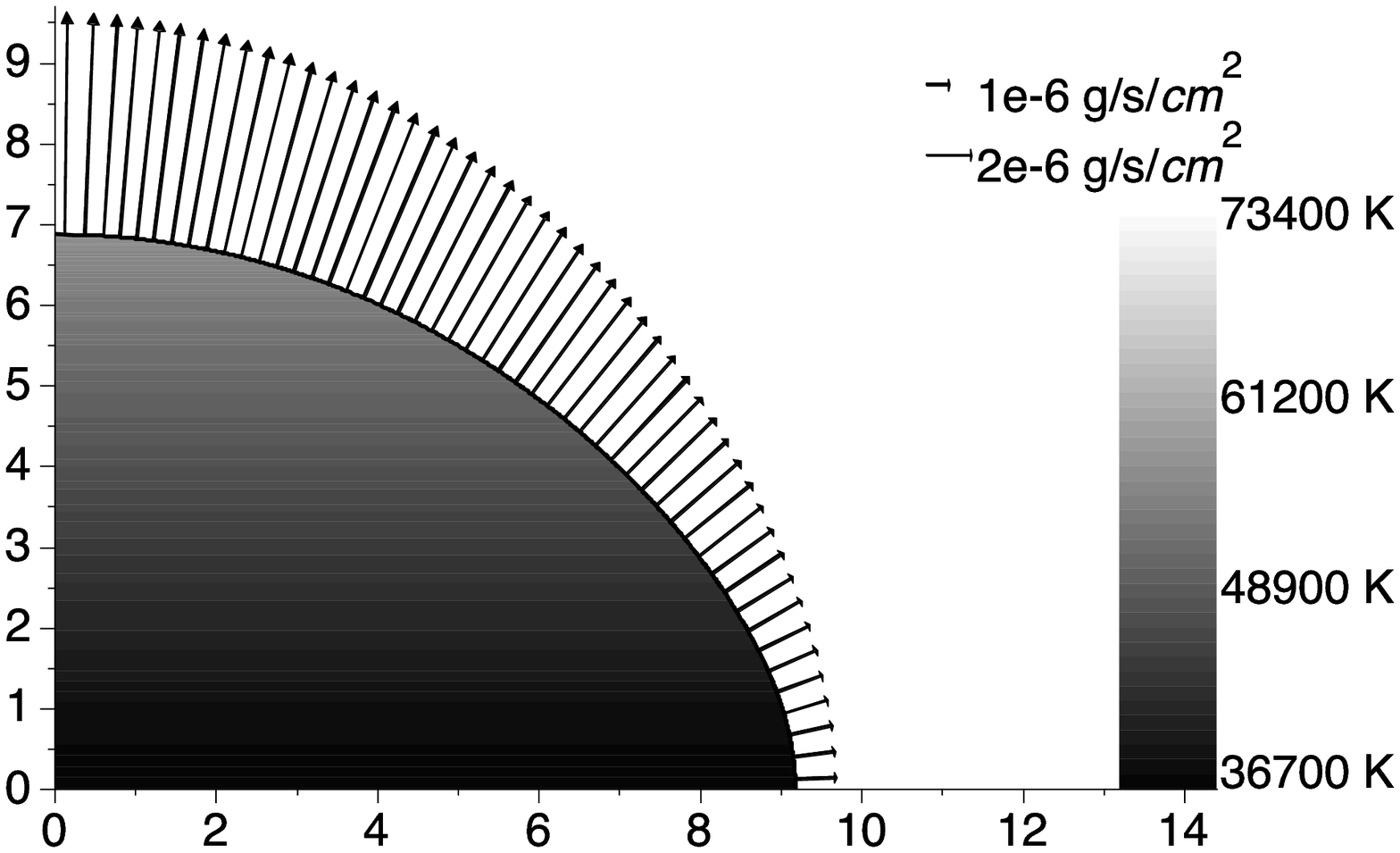}
\hfill
\includegraphics[scale=0.32,angle=0]{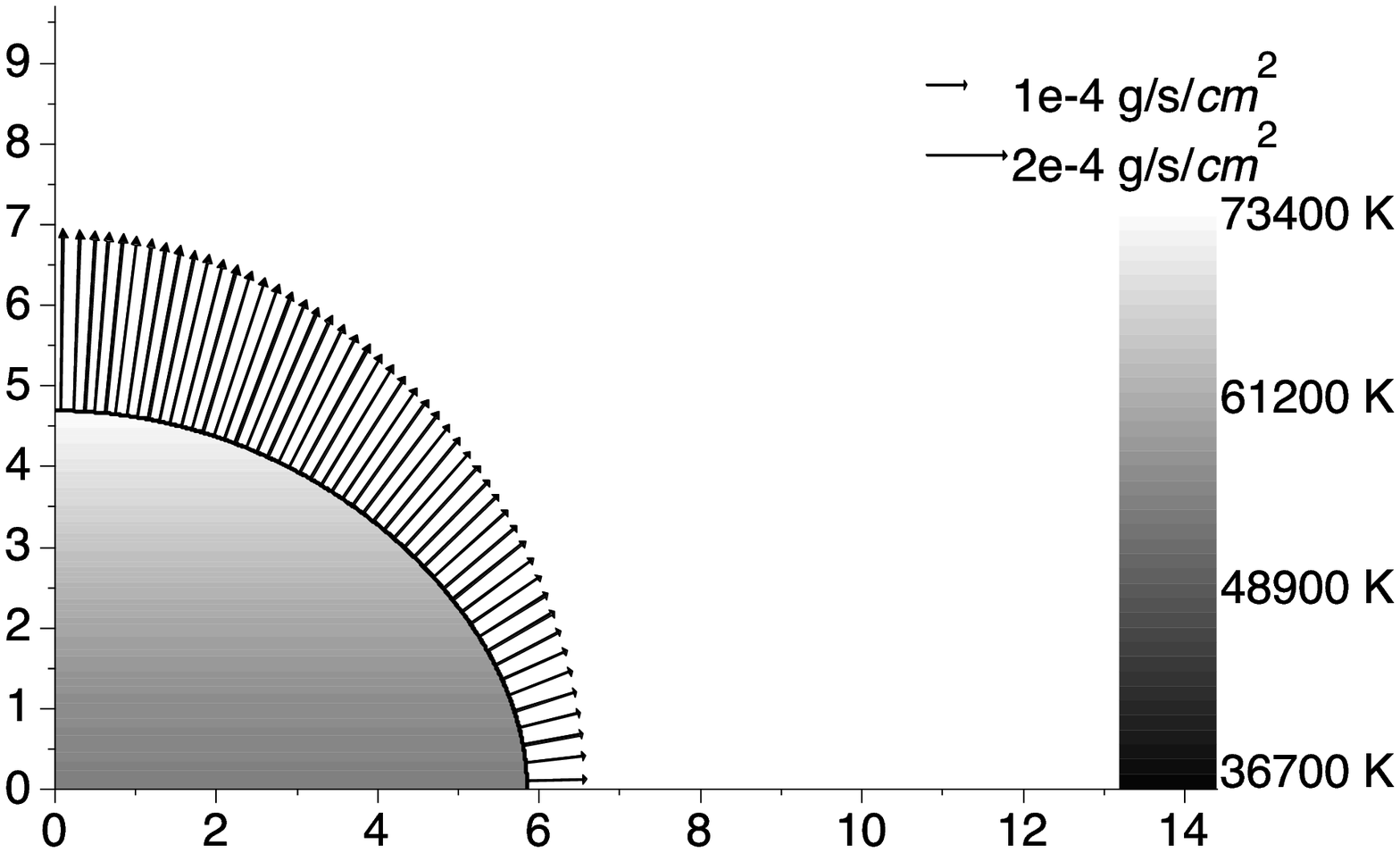}
\caption{{\it Left panel}: Variation of the mass flux at the surface of an initial 35M$_\odot$ 
at a stage during the core H-burning phase when the mass fraction of hydrogen at the centre is 
$X_c=0.42$. The axis are in units of solar radius.
The velocity of the star on the ZAMS is 550 km s$^{-1}$ corresponding to $\Omega/\Omega_{\rm crit}=0.84$. The star follows a homogeneous evolution. At the stage represented  
$\Omega/\Omega_{\rm crit}\sim 1$. 
{\it Right panel}: Same as the left panel, but for a later stage with $X_c=0.02$ and  
$\Omega/\Omega_{\rm crit}\sim 1$. Courtesy of C. Georgy.
}\label{aniso1}
\end{figure}


\section{The WC/WN number ratios}

The variation with metallicity of the number of WC to WN stars has been often discussed in this meeting. It is a well known fact that this ratio increases with the metallicity (see Fig.~\ref{tma} left panel\footnote{we consider here regions having reached a stationary situation, {\it i.e.} regions where the star formation rate can be considered to have remained constant for the last twenty million years.}). 
Many attempts have been performed to reproduce the observed trend: for instance the enhanced mass loss rate models of Meynet \& Maeder (1994) provided a good agreement for solar and higher than solar metallicity but produced too few WN stars in metal-poor regions. The inclusion of rotation together with reduced mass loss rates accounting for the effects of clumping improved the situation in the metal poor region, but produced too many WN stars at solar and higher metallicities
(Meynet \& Maeder 2005).
Eldridge \& Vink (2006) show that  models that include the mass-loss metallicity scaling during the WR phase closely reproduce the observed decrease of the relative population of WC over WN stars at low metallicities. However such models severely underestimate the fraction of WR to O-type stars. In that case, to improve the situation, a high proportion of
Wolf-Rayet stars  originating from mass transfer in close binaries have to be assumed at all metallicities. For instance at solar metallicity about 75\% of the WR stars should be produced in close binary systems (see Fig.~5 in Eldridge et al. 2008). 

Recently we reexamined this question starting from our rotating stellar models (Meynet et al. in preparation). First let us recall that
in Meynet \& Maeder (2005, 2006), the most massive rotating stars enter into the WR regime already during the MS phase. This feature has good and bad effects. On one hand, it allows these models to well reproduce the variation
of the number fraction of WR to O-type stars since it significantly increases the WR lifetimes. On the other hand, it produces very long WN phases since the star enters into the WR phase having still a huge H-rich envelope. As a consequence,
too low values for the WC/WN ratio are obtained at solar and higher metallicities.

\begin{figure}
\includegraphics[width=2.6in,height=2.0in,angle=0]{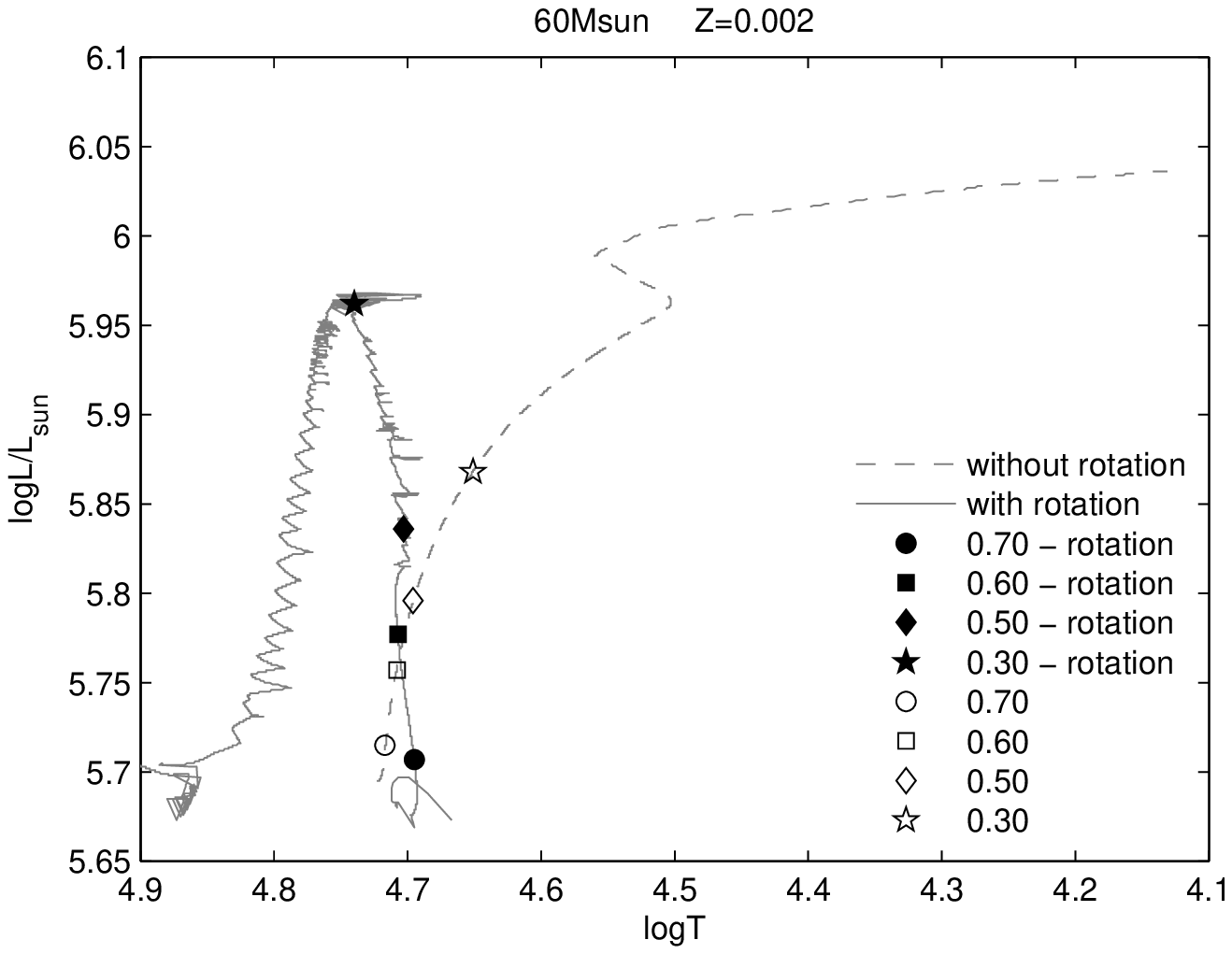}
\hfill
\includegraphics[width=2.6in,height=2.0in,angle=0]{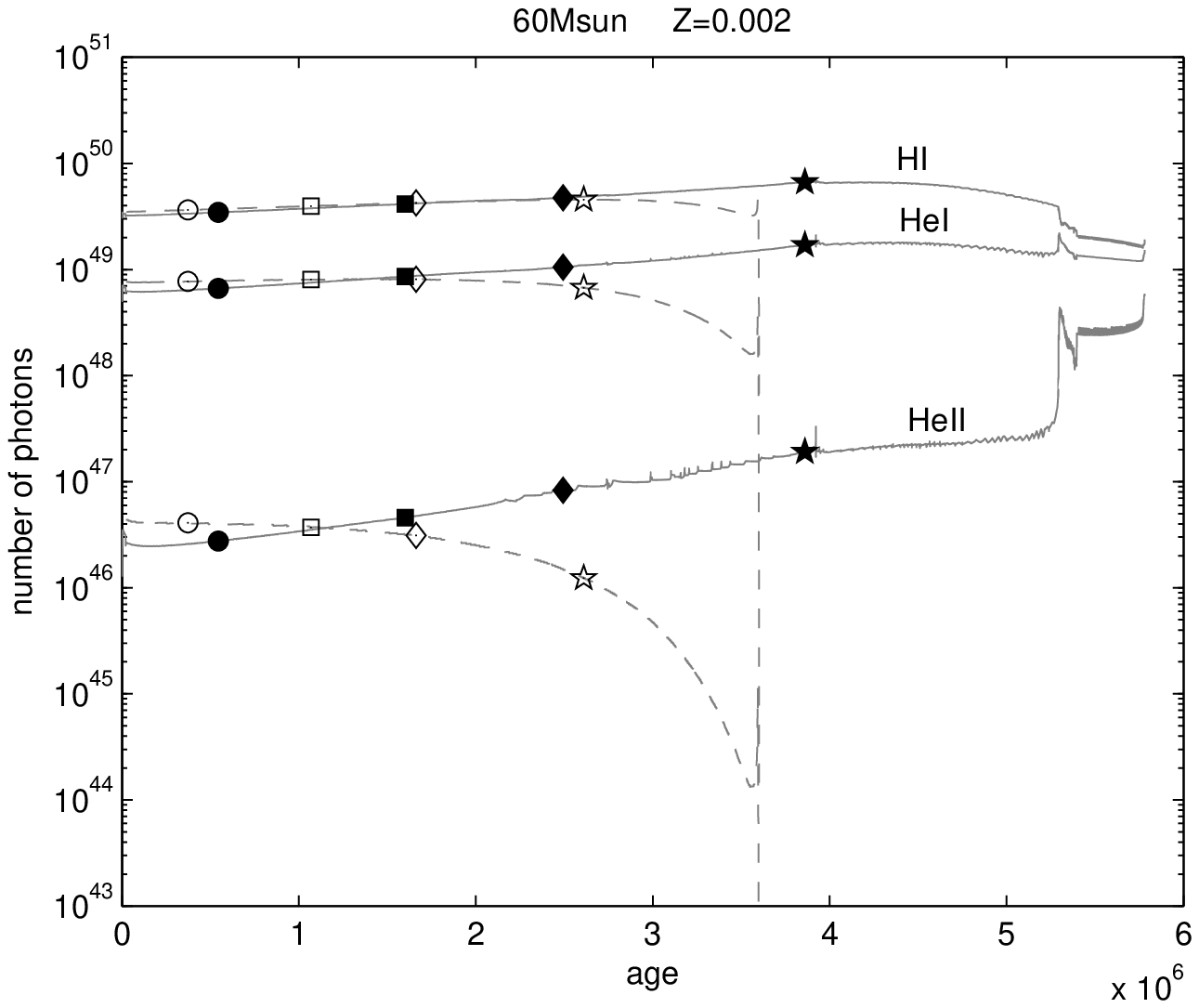}
\caption{{\it Left panel}: Evolutionary tracks in the HR diagram for a rotating 
(continuous line) and a non-rotating 60M$_\odot$ stellar model (dashed line)
at $Z=0.002$. The rotating model has an initial velocity of 530 km s$^{-1}$
or $\Omega/\Omega_{\rm crit}=0.75$. Points indicate the position of the model when
various values of the hydrogen mass fraction at the centre are reached.
{\it Right panel}: Evolution as a function of the age of the ionising luminosity.
Continuous lines are for the rotating model, dashed lines are for the non-rotating ones.
Courtesy of C. Beffa.
}\label{reio}
\end{figure}

In the above computations, we made the hypothesis that 
when a star enters into the WR stage during the MS phase,  it
avoids the Luminous Blue Variable phase. This is probably not correct. A more realistic solution is to consider that a star which becomes a WR star during the MS phase, enters a LBV phase after
the core H-burning phase, before evolving back into the WR regime. 
When this more realistic solution is applied, reasonable values for both the WR/O and the WC/WN ratios are obtained.
Indeed the ratios of WR/O and of WC/WN given by these models at the solar metallicity are
0.06 and 0.9 which compare reasonably well with the observed values of 0.1 and 0.9 respectively.
Both ratios are not reproduced by the non-rotating models to which a similar solution is applied.
At the moment only the case at solar metallicity has been computed, but we are confident that such a scenario will also provide reasonable answers at other than solar metallicities. 
This discussion illustrates the possible key role that the LBV phase may play in shaping the
WC/WN ratio. 

\section{The changes of the surface abundances}
\label{secabo}
 
Models without rotation predict no surface enrichments before the red supergiant stage for stars less massive than about 40M$_\odot$. This is clearly in contradiction with the observations.
In contrast,  
rotationally induced mixing produce changes of the surface abundances already during the MS phase. This is shown
in Figs.~\ref{SMC920} and \ref{LMC13SMC15}. Model results depend on the 
physics included and on the choice of some parameters. In Figs.~\ref{SMC920} left and \ref{LMC13SMC15} right, two series of models are plotted: the models of Ekstr\"om et al. (2008) have been computed with the same physics as in Meynet \& Maeder (2005), {\it i.e.} the expression for the horizontal turbulence was taken as proposed in Zahn (1992), the value of $\alpha$ in the expression of the shear diffusion 
coefficient was taken equal to 2 (see Eq.~3 in Maeder \& Meynet 2001), and an overshooting parameter of 0.1$H_p$ was used (these models will be called models A hereafter),
the models B by Ekstr\"om et al.
(in preparation) differ in many aspects. We just mention the two most important ones: they use
the expression for the horizontal turbulence by Maeder (2003) and the value of $\alpha$ in the expression of the shear diffusion 
coefficient was taken equal to 4. We can clearly
see that, given an initial value of $\Omega_{\rm ini}/\Omega_{\rm crit}$,
models B are more efficiently mixed than models A. 
One sees also that models A, even spanning the whole range
of possible $\Omega_{\rm ini}/\Omega_{\rm crit}$ values cannot reproduce the highest enrichments.
Models B in contrast do appear in better position to reproduce the observed range of values and are thus to be preferred. 

In Fig.~\ref{SMC920} right, rotating models computed with the same physics as models B have been computed for the LMC metallicity (long-dashed lines). Tracks computed with the effects of a magnetic field as in Meynet \& Maeder (2007) are also shown for a few velocities (continuous lines).
We see that models with magnetic fields are more efficiently mixed, they also show a more progressive surface enrichment. Both series of models can reproduce the observed range of values. One notes however that models with magnetic fields appear in a better position for explaining the most extreme cases.
In Fig.~\ref{LMC13SMC15} left, the models B (long-dashed lines) are plotted together with the models of Maeder \& Meynet (2001). Again models B are more efficiently mixed and appear in better position for reproducing the observed enrichments. 

On the whole, we see that the recent observations obtained in the large surveys described above support quite efficient mixing processes. The models of Ekstr\"om et al. (in preparation)
can account for a great part of the observed range. More detailed comparisons will be presented elsewhere.   

In Fig.~\ref{tma} right, the continuous lines show the evolutions of the surface N-enrichment as a function of the He-enrichment at the surface of different models. 
The dotted line shows the 
evolution in the convective core of a 20M$_\odot$ stellar model with $Z=0.002$. The evolution
in the convective core represents an upper envelope of what we can expect from a theoretical point of view. Note that this curve does not much depend on the initial mass, but depends on the initial CNO content as can be seen by comparing the dotted and the dashed line. The dashed line is obtained from models with initial abundances for the SMC as given by
\cite[Hunter et al. (2007)]{Hunter_etal07}, while the dotted one is obtained starting with
the relative abundances as given by \cite[Asplund et al. (2005)]{AS05}.

Let us emphasize that progression along the ``surface'' (continuous lines) and ``core'' (dotted) lines goes at a different pace. This can be realized by noting that on the ``surface'' line the end
of the MS phase in general occurs very early (typically the end of the MS is indicated by a star on the track 2, corresponding to the same model as the one used to draw the dotted line), while, on the ``core'' line, the end of the MS phase would occurred at $Y_s-Y_c=0.75$ and
$\Delta$log N/H tending toward infinity since the hydrogen abundance is zero.
The N-enrichment occurs very rapidly at the surface well before any surface helium enrichment. This is of course due to the fact that very rapidly (see the dotted line) the nitrogen abundance increases at the centre creating thus a strong chemical gradient between the core and the envelope.
Since diffusive velocity is greater when the gradient of abundance is greater, the presence of such a strong gradient favors a rapid mixing. 
The gradient in helium is much more shallow and makes the diffusion of this element to occur on much longer timescales.

With these remarks in mind, let us now discuss each model: 
model 1 shows the behavior of a fast rotating 60 M$_\odot$ model at $Z=0.002$ following a nearly homogeneous evolution. The evolution of its surface abundances approaches that of the central ones.
The whole portion of the track shown in Fig.~\ref{tma} right occurs during the MS phase. Such a model would easily account for the observed He enrichments during the MS phase.
The model 2 (20 M$_\odot$ with
$v_{\rm ini}=304$ km s$^{-1}$ and for $Z=0.002$) shows at the end of the MS phase an increase
in helium of 0.01 (in mass fraction) and an increase of 0.7 dex in N/H. The He-enrichment
is lower than the observed one.
Models 3 and 4 compare two 13 M$_\odot$ models for the LMC composition. Both models have very high initial rotation (382 km s$^{-1}$) but model 3 is computed without magnetic field while model 4 is computed with a magnetic field. We see that despite the high velocity, the model without magnetic field does not have any surface helium enrichment. Model with magnetic field reaches at the end of the MS phase an enrichment of nearly 0.015 with respect to its initial value. This enrichment is however too low to account for the He-enrichments obtained by \cite[Huang \& Gies (2006b)]{HuangGies06b}
at the end of the MS phase. 

We conclude from these comparisons that the present-day estimates of He-enrichments at the surface of some stars require tracks following a nearly homogeneous evolution\footnote{Note that many stars show N enrichments but no He enrichment, they would stand on the vertical line at the abscissa 0 in Fig.~\ref{tma} right. These stars can in general be explained by models with normal rotation velocities.}. 


\section{Wind anisotropies and homogeneous evolution}

Rotation has a deep impact on the way massive stars are losing mass (see the papers by A. Maeder and
R. Hirschi in these proceedings). Here
we focus on the anisotropies of the winds induced by fast rotation (Maeder 1999).
When the surface velocity is near the critical one strong wind anisotropies appear. This is illustrated in Figs.~\ref{aniso1} which shows the variation of the mass flux at the surface of a fast rotating 35 M$_\odot$ model following a homogeneous evolution.
The lengths of the arrows are proportional to
the mass flux. 
Figure~\ref{aniso1} left shows a stage when the star is an O-type star, the right panel when it is a WR star.  
The scale for the mass flux was changed between left and right panels
since during the WR phase
the mass fluxes are about 2 orders of magnitudes higher.
The grey scale varies as a function of the effective temperature (von Zeiple effect).
 
Accounting for wind anisotropies allows to keep in the star about
25-30\% of the angular momentum which would be lost by an isotropic wind. 
This difference
produces important differences in the way the angular momentum is distributed 
in the star at the pre-supernova stage. Typically, 
when wind anisotropies are considered, the specific angular momentum in the 3 inner
solar masses are more than  a factor 6 higher than when the isotropic winds are
considered (see Meynet \& Maeder 2007).
Thus wind anisotropies are probably a key feature of homogeneous evolutionary tracks
and of GRB progenitors.

Such homogeneous evolution will also produce higher
outputs of ionising photons as can be seen in Fig.~\ref{reio}.
The total number of photons released during the whole stellar lifetime
with an energy sufficient to ionize H
is about a factor 2 greater (passing from 1.46 10$^{56}$ to 2.65 10$^{56}$)
for the blueward track than for the normal one.
Most of these photons (98\%) are emitted during
the core H-burning phase.
The total number of HeI ionising photons
passes from 2.4 10$^{55}$ to 6.7 10$^{55}$.
Most of these photons are released during the
core H-burning phase (92\%).
The number of HeII ionising photons
passes from 
8.7 10$^{52}$ to 184 10$^{52}$.
In that case 2/3 are emitted during
the core He burning phase. 
Thus homogeneous evolution increases the number
of HeI and HeII photons 
by about a factor 3
and 20 respectively.
Therefore, if homogeneous evolution is
a not too rare scenario for PopIII and very metal poor massive stars, 
then the budget of ionising photons in the early Universe should take into
account such sources.

\section{Conclusions}

Considering a massive star of given initial mass
and metallicity, one encounters for increasing angular momentum content the following
effects of rotation:
The first effect of rotation which already occurs
for modest rotation rate is internal mixing. Of course
when the velocity increases the mixing also increases.
Then increasing the angular momentum content, the 
second series of effects induced by rotation concerns mass loss.
Rotation may trigger mechanical mass loss, it may also
at very low metallicity increase the ``metallicity''
of the outer layers (metals produces by the star itself)
and thus increase the opacity and trigger radiative stellar winds
(see the papers by Hirschi and Ekstr\"om in the present volume).
Finally as an extreme case of the effects of rotation, there are
the homogeneous evolution. Theory indicates that
many of these effects vary as a function of Z, tending to be more
important in metal poor regions. At the moment many indirect observational
features seem to require high rotation rates at low Z for massive stars
(see Meynet et al. 2008).
Direct observations, limited to MC metallicities, seem also to support this view by
indicating that rotation rates appear to be higher at lower Z.
Therefore rotation is probably
a key effect for understanding the first generations of massive stars.

\begin{discussion}

\discuss{Koenigsberger}{Could you say a few words on how the gradient of mean molecular weight can affect the efficiency of rotational mixing?}

\discuss{Meynet}{When the molecular weight increases with depth as is the case in stars, mixing becomes more difficult, since greater energy is required to lift off heavy material and to mix down light one. This can be shown through the expression of the Richardson criterion (see Maeder \& Meynet 1996, A\&A, 313, 140).
Meynet and Maeder (1997, A\&A, 321, 165) have shown that the strict application of the Richardson criterion would prevent any mixing in regions with a molecular weight gradient. Only when the effects of the strong horizontal turbulence are accounted for in a proper way (Talon \& Zahn 1997,
A\&A, 317, 749) can mixing be efficient enough.}

\discuss{Limongi}{A comment on the anticorrelations observed in globular clusters. In principle AGB models can explain the anticorrelations maintaining the C+N+O=const. In the most massive AGB stars, HBB occurs and also if there are few dredge-up episodes the chemical composition observed in the turn off stars of globular clusters can be reproduced. the problem is that to quantitatively account for the observations a very peculiar IMF is required. Now a question concerning the WC/WN ratio: how much the WN/WC problem depends on the mass loss rate during the red supergiant phase?}

\discuss{Meynet}{The WR/O and the WC/WN ratios can be reproduced without modifying the mass loss rates in 
the red supergiant phase. Thus, the mass loss rate during 
the RSG does not appear as a critical point. However
higher mass losses during the red supergiant phase might allow lower initial 
masses to enter the WN regime and thus may have 
an impact on this ratio. At the moment however, except may be in the 
clusters near the galactic centre, there is no single-aged clusters 
showing both a significant red supergiant and Wolf-Rayet population. 
This indicates that the two populations originate from different initial mass ranges.}

\discuss{Hunter}{While fast rotation does change the surface gravities, it does not change the age of all the stars and hence the bulk of the ``older'' fast rotating unenriched stars remain. Secondly, given the rotational velocity distribution, the low $v\sin i$ stars are not due to $\sin i$ effects, but are indeed slowly rotating stars.}

\discuss{Aerts}{1) The study of Morel et al. (2006) contains stars with N-enrichment and a seismic estimate of the equatorial rotational velocity, so here one is not bothered with $\sin i$ uncertainty. 2) The FLAMES survey contains two clusters with a lot of $\beta$Cephei pulsators, so a line-profile analysis of these stars can also lead to $\sin i$. this could help to interpret the ``Hunter-diagram''.} 

\end{discussion}

\end{document}